\documentclass[twocolumn,prb,showpacs,showkeys]{revtex4}

\usepackage{graphicx}
\usepackage{dcolumn}
\usepackage{bm}

\begin{document}

\title{{Mirror Nesting of the Fermi Contour and \\
Superconducting Pairing from the Repulsive Interaction}}

\author{V.~I.~Belyavsky}
\altaffiliation {State Pedagogical University, Voronezh, 394043,
Russia} 
\author{Yu.~V.~Kopaev}
\email{kopaev@sci.lebedev.ru}
\author{S.~V.~Shevtsov}
\altaffiliation{State Pedagogical University, Voronezh, 394043,
Russia} \affiliation{Lebedev Physical Institute, Russian Academy
of Sciences, Moscow, 119991, Russia}

\begin{abstract}
We consider the necessary conditions of superconducting pairing at
repulsive interaction between particles composing a pair with
large total momentum: (1) the existence of, at least, one negative
eigenvalue of the repulsion potential and (2) mirror nesting of
the Fermi contour. Under these conditions, we represent the
solution of the self-consistency equation continuously depending
on the momentum of the relative motion of the pair. The
corresponding superconducting order parameter changes its sign on
a line crossing the Fermi contour inside the domain of definition
of the relative motion pair momentum. We argue that
repulsive-induced superconducting pairing with large total pair
momentum may be just the case relating to high-temperature
superconducting cuprates.

\end{abstract}

\pacs {78.47.+p, 78.66.-w}

\keywords{superconductivity, repulsion interaction, nesting, large
pair momentum, cuprates}

\maketitle

\section{Introduction}

A conventional point of view relating to the mechanism of
high-temperature superconductivity of cuprates is absent up to
this point. Nevertheless, there is an experimental evidence that
there is the {\it {singlet}} pairing of current carriers below the
superconducting (SC) transition temperature in these compounds
although the question is what interaction dominates pairing.
\cite{Scalapino} In this connection, the models of the
superconductivity both with dominating attraction and repulsion
are considered. \cite{Aoki} Another open question is what
contribution into the pair binding energy (due to potential energy
as it is in the conventional theory by Bardeen, Cooper and
Schrieffer (BCS) \cite{BCS} or, maybe, due to kinetic energy
\cite{Hirsch}) may be considered as a driven force of the SC
transition in the cuprates. \cite{Norman}

In contrast to the conventional paradigm that the SC transition is
a consequence of an instability of the Fermi liquid with respect
to arbitrarily weak attraction between particles the question
relating to the character of the ground state of normal (N) phase
of cuprate compounds remains under discussion. \cite{Orenstein}
The manifestation of the pseudogap in underdoped cuprates is a
reason to suppose that the N phase may be something different from
the normal Fermi liquid. Assuming that cuprates are strong
correlated electron systems one can, in principle, connect rather
wide spectrum of physical properties of these compounds which
include, besides typical of cuprates superconductivity and
antiferromagnetism (AF), some other ordered states arising due to
a competition of SC and AF orders. \cite{Chakravarty} In this
connection one can raise the question of the applicability of the
mean-field theory (like the BCS theory which describes
conventional superconductors successfully) to cuprate
superconductors. \cite{Sachdev}

Pairing with zero total pair momentum does not impose any
kinematical constraint on the momenta of the relative motion
coinciding, in this case, with the momenta of the particles
composing the pair. In the BCS model with an attractive
interaction between particles, however, there is a ``dynamical''
constraint connected with the fact that the domain of the
effective attraction is a narrow layer enveloping the Fermi
surface. The volume of this layer is proportional to the
statistical weight of the pair that is a number of one-particle
states which contribute to the state of the relative motion of the
pair. \cite{Schrieffer}

On the contrary, at non-zero total pair momentum there is an
essential kinematical constraint resulting in a finite domain of
the momentum space in which momenta of the relative motion of the
pair should be defined.\cite{Fulde} A supplementary dynamical
constraint arising from an attractive interaction, decreases such
a domain abruptly resulting in a considerable decrease in the
statistical weight of the pair. It should be emphasized that, in
the case of Coulomb repulsion, any dynamical constraint is absent
and the statistical weight of the pair is fully determined by the
volume of kinematically allowed domain of the momentum space.

The Fermi surface of quasi-two-dimensional (2D) electron system of
copper--oxigen planes of a cuprate compound is degenerated into a
line which we call the Fermi contour (FC). In cuprates, the FC is
situated in an extended vicinity of a saddle point of the electron
dispersion. \cite{Shen} As a consequence, one may expect a FC
with strong nesting feature and, correspondingly, a considerable
difference in the transversal (with respect to the FC) and the
longitudinal Fermi velocities. \cite{Chiao} In addition, one can
assume that the corresponding effective masses should be of
opposite signs and differ in absolute values considerably. Thus,
one can expect that, for some definite (antinodal) directions
coinciding with the sides of 2D Brillouin zone, there exists
rather large area of 2D momentum space defining the statistical
weight of the pair with certain non-zero total momentum.
\cite{BKK_1} One can create a current-less state corresponding to
such a pair as a definite linear combination of the states
relating to crystal equivalent momenta. \cite{BKK_1}

The SC instability of 2D electron system arises due to the fact
that the boundary separating occupied and unoccupied one-particle
states is a line. In the case of zero total pair momentum, this
line is the full FC whereas at non-zero total pair momentum such a
boundary is, generally speaking, a set of points. Therefore, in
such a case, there is no logarithmic singularity in the
self-consistency equation and, consequently, the SC pairing is
impossible at arbitrarily weak interaction strength.

However, the extended saddle point may result in a rise of such a
topology of isolines in a vicinity of the FC that the boundary
between occupied and unoccupied states of the pair relative motion
turns out to be a set of lines of finite length. These lines form
a ``pair'' Fermi contour (PFC) for the relative motion of the pair
with definite total momentum. Such a feature ({\it {mirror
nesting}}) of the FC \cite{BK_1} is held in a finite doping
interval and may warrant a logarithmic singularity in the
self-consistency equation resulting in the SC instability at
arbitrarily interaction strength.

The channel of the SC pairing with large total pair momentum
results quite naturally from a consideration of the competition of
Cooper (at zero pair momentum) SC pairing and AF ordering in the
framework both a band model \cite{Kopaev} and the models which
concern strong electron correlations. \cite{Zhang} Also, such a
channel arises with necessity in rather general phemenology based
on symmetry considerations. \cite{Guidry}

Recent experimental results relating to neutron scattering in
cuprates (the so-called 41 meV peak) may be considered as an
indirect evidence in favor of the SC pairing with large total
momentum. \cite{Zhang} Another two fundamental experimental
results, the so-called ``peak--dip--hump structure'' of
angle-resolved photoemission spectrum (ARPES) \cite{Campuzano,
Damascelli} and relatively small superfluid phase stiffness
typical of cuprates, \cite{Orenstein} can be explaned
qualitatively in the framework of {\it {the conception of mirror
nesting and repulsion-induced SC pairing with large total pair
momentum }} as well.

The goal of this paper is an investigation of the necessary
conditions of the repulsion-induced SC pairing with large total
pair momentum and a solution of the corresponding self-consistency
equation arising within the mean-field scheme.

\section{Mirror nesting}

The kinematical constraint due to the presence of the FC results
in the fact that both momenta $\bm{k}_{\pm}^{}$ of like-charged
particles composing a pair with total momentum $\bm{K}=\bm{k}_+^{}
+ \bm{k}_-^{}$ have to be situated either inside or outside the
FC. Thus they belong to dependent on $\bm{K}$ domain of the
momentum space which may be considered as a statistical weight of
the pair. This domain ${\Xi}^{}_K$ is symmetrical with respect to
the inversion transformation $\bm{k}\leftrightarrow -\bm{k}$ of
the momentum of the relative motion of the pair,
$\bm{k}=(\bm{k}_+^{}-\bm{k}_-^{})/2$, and, generally speaking,
consists of the two parts $\Xi_K^{(-)}$ and $\Xi_K^{(+)}$, inside
and outside the FC, respectively. An excitation energy of the pair
with respect to the chemical potential $\mu$ that is kinetic
energy of the particles composing the pair,
\begin{equation}
\label{2_01} 2\xi_{Kk}^{}={\varepsilon}
({\bm{K}}/2+{\bm{k}})+{\varepsilon} ({\bm{K}}/2-{\bm{k}})-2\mu,
\end{equation}
equals zero on the boundary separating occupied ($\Xi_K^{(-)}$)
and unoccupied ($\Xi_K^{(+)}$) parts of the domain ${\Xi}^{}_K$.
Hereafter, for the sake of simplicity, we restrict ourselves to a
consideration of hole-doped cuprate superconductors within a
simple single-band scheme; therefore, ${\varepsilon}({\bm
{k}}_\pm^{})$ makes sense of a hole dispersion.

As it is mentioned above, at $K= 0$, the momenta of the relative
motion ${\pm {\bm k}}$ coincide with the momenta of the holes
composing the pair and therefore the boundary separating the
occupied and unoccupied parts of the domain ${\Xi}^{}_K$ is the
full FC whereas at any $K\neq 0$ the boundary between
$\Xi_K^{(-)}$ and $\Xi_K^{(+)}$, generally speaking, turns out to
be a set of isolated points. However, due to a special hole
dispersion, such a boundary, at definite total pair momentum, may
coincide with finite-length pieces of the FC as it is shown
schematically in Fig.1. In such a case the boundary between
$\Xi_K^{(-)}$ and $\Xi_K^{(+)}$ plays role of a peculiar FC with
respect to the relative motion of the pair and may be called a
{\it {pair Fermi contour}} (PFC).

Using the symmetry property, $\bm{k}\leftrightarrow -\bm{k}$, one
can write the condition defining the PFC in the form
\begin{equation}
{\varepsilon}({\bm{K}}/2+{\bm{k}})-\mu={\varepsilon}
({\bm{K}}/2-{\bm{k}})-\mu,\label{2_02}
\end{equation}
under the additional condition that both $\bm{k}$ and $ -\bm{k}$
belong to the domain ${\Xi}^{}_K$. It is obvious that, to obey the
condition (\ref{2_02}), the vector $\bm{K}$ should be directed
along one of the symmetrical (antinodal or nodal) direction of 2D
Brillouin zone. Hence the boundary separating occupied and vacant
states has to obey a mirror symmetry with respect to the pair
momentum $\bm{K}$ direction and so Eq.(\ref{2_02}) may be called
{\it {a mirror nesting}} condition. It should be also noted that,
under mirror nesting condition, the inversion transformation
$\bm{k}\leftrightarrow -\bm{k}$ superposes an occupied state with
an occupied one and a vacant state with a vacant one,
respectively.

\begin{figure}
\includegraphics[]{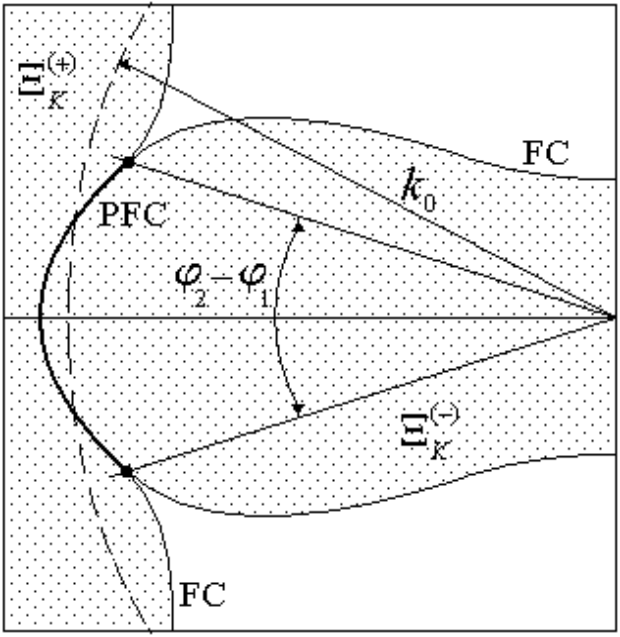}
\caption[*]{\label{bks_01.eps} Domain of definition ${\Xi}^{}_K
={\Xi}_{K}^{(-)}+{\Xi}_{K}^{(+)}$ of pair relative motion momentum
(schematically; it is shown only a half of the domain): the
subdomain ${\Xi}_{K}^{(-)}$ (${\Xi}_{K}^{(+)}$) corresponds to
the shaded occupied (unoccupied) part of the domain $\Xi$. These
subdomains are separated by the PFC (solid line) being a part of
the FC (fine line). The nodal line $k=k^{}_0$ (dashed line)
separates the parts of the momentum space corresponding to
different signs of the order parameter.}
\end{figure}

In the framework of a single-band scheme relating to cuprates, one
can give some rather obvious examples of the mirror nesting
condition resulting in a rise of the PFC. \cite{BK_2, BKK_2} In a
case when the FC of a hole-doped cuprate compound is simply
connected and appears as a square with rounded corners,
\cite{Shen} one can expect that a change of FC curvature sign,
which can be easily realized because of the existence of long,
almost rectilinear, pieces of the FC, may provide a realization of
the mirror nesting condition for certain chosen pair momenta.
\cite{BKK_2} One more single-band example represents an underdoped
cuprate compound considering as doped Mott insulator in which the
FC encloses some symmetrically disposed hole pockets around the
points belonging to symmetrical directions in 2D Brillouin zone.
\cite{Hlubina} In such a case, the mirror nesting condition is
perfectly satisfied when a half of the total pair momentum
corresponds to the center of a pocket and the whole of the line
enclosing the pocket turns out to be the PFC. The stripe structure
arising in underdoped cuprates gives one more possibility
resulting in a rise of PFC. \cite{BK_2}

\section{The self--consistency equation}

To consider the problem of pairing with non-zero total pair
momentum in the mean-field framework let us write the Hamiltonian
corresponding to pair relative motion in the form
\begin{eqnarray}\label{3_01}
\hat H_K = && \sum_k {\left [ (\varepsilon^{}_{{\bm k}_+}-{\mu})
{\hat{a}^{\dag}_{k_+\uparrow}}{\hat{a}^{}_{k_+\uparrow}}
 + (\varepsilon^{}_{{\bm k}_-}-\mu) {\hat{a}^{\dag}_{k_-\downarrow}}
{\hat{a}^{}_{k_-\downarrow}}\right ]}\nonumber\\
&& +{\frac 1{S}} \sum_{k,k^\prime} U({\bm k}- {\bm k^\prime})
{\hat{a}^{\dag}_{k_+\uparrow}} {\hat{a}^{\dag}_{k_-\downarrow}}
{\hat{a}^{}_{{{k^\prime}_-}\downarrow}}
{\hat{a}^{}_{{{k^\prime}_+}\uparrow}}
\end{eqnarray}
where $\varepsilon^{}_{{\bm k}_{\pm}}\equiv \varepsilon ({{\bm
k}_{\pm}})$, $U({\bm k} - {\bm k^\prime})$ is the Fourier
transform of the interaction energy, $S$ is a normalizing area,
${\hat a}_{k_{\pm}\sigma}^{\dag}$ (${\hat a}_{k_{\pm}\sigma}^{}$)
creates (annihilates) a hole with the momentum ${\bm k}_{\pm}$ and
the spin quantum number $\sigma =\; \uparrow ,\downarrow$. The
summation in Eq.(\ref{3_01}) is taken over all range of the
allowed values of the momentum of the relative motion and thus is
restricted by the domain $\Xi_K$. Note that the summation in the
Hamiltonian Eq.(\ref{3_01}) should be taken over only two (instead
of three in a general case) variables, $\bm k$ and ${\bm
k}^\prime$, just as in the case of BCS Hamiltonian, taking into
account the only interaction between particles composing the pairs
with total pair momentum ${\bm K}$.

As usual, to diagonalize the Hamiltonian Eq.(\ref{3_01})
approximately one can introduce creation and annihilation
operators of new quasiparticles using the well-known
Bogoliubov--Valatin transformation:
\begin{eqnarray}
\label{3_02} {\hat{a}_{k_+\uparrow}^{}}=u_{Kk}^{}
{\hat{b}_{k,{+1}}^{}}+ v_{Kk}^{} {\hat{b}^{\dag}_{k,{-1}}} ,\nonumber\\
{\hat{a}_{k_-\downarrow}^{}}=u_{Kk}^{} {\hat{b}_{k,{-1}}^{}}-
v_{Kk}^{} {\hat{b}^{\dag}_{k,{+1}}}.
\end{eqnarray}
The Hamiltonian, up to the terms of the order of ${\hat b}^2$, can
be written as
\begin{equation}
\label{3_03}{\hat H}_K^{} =E_{K0}^{} + {\hat H}_K^{(0)} + {\hat
  H}_K^{(1)}.
\end{equation}
The ground state energy has the form
\begin{equation}
\label{3_04}E_{K0}^{}=2\sum_k \xi_{Kk}^{}v^2_{Kk} + \sum_k
\Delta_{Kk}^{} u_{Kk}^{}v_{Kk}^{},
\end{equation}
where, related to the value of the chemical potential, an energy
of the relative motion of the pair with the total momentum ${\bm
K}$ is defined by the Eq.(\ref{2_01}) and the order parameter is
defined as
\begin{equation}
\label{3_05}\Delta_{Kk}^{} = {\frac 1S} \sum_{k^\prime} U({\bm
k}-{\bm k^\prime}) u_{Kk^\prime}^{} v_{Kk^\prime}^{}.
\end{equation}
Diagonal and nondiagonal, with respect to quasiparticle operators,
parts of the Hamiltonian can be written as
\begin{equation}
\label{3_06}\hat{H}_K^{(0)} = \sum_{k;\beta =\pm 1}
\eta_{K\beta}^{} ({\bm k})
{\hat{b}^{\dag}_{k,{\beta}}}{\hat{b}_{k,{\beta}}^{}}
\end{equation}
and
\begin{eqnarray}
\label{3_07}\hat{H}_K^{(1)}= && \sum_k \left [2\xi_{Kk}^{}
u_{Kk}^{}v_{Kk}^{} -
(v^2_{Kk} - u^2_{Kk})\Delta_{Kk}^{} \right ] \nonumber\\
&& \times ({\hat{b}^{\dag}_{k,{+1}}}{\hat{b}^{\dag}_{k,{-1}}} +
{\hat{b}_{k,{-1}}^{}}{\hat{b}_{k,{+1}}}^{}),
\end{eqnarray}
respectively. Here the energies corresponding to two branches
($\beta =\pm1$ ) of the one-particle excitation spectrum are equal
to each other,
\begin{equation}
\label{3_08}\eta_{K\beta}^{}({\bm k}) = \sqrt {\xi^2_{Kk} +
\Delta^2_{Kk}}.
\end{equation}

A choice of the amplitudes, $u_{Kk}^{}$ and $v_{Kk}^{}$, in
Bogoliubov--Valatin transformation Eq.(\ref{3_02}) is determined
in the zero-temperature limit by the conditions that (i) all of
the states inside the subdomain $\Xi_K^{(-)}$, in which the
kinetic energy of the relative motion of the pair is negative,
$2\xi_{Kk} < 0$, must be occupied and (ii) the nondiagonal part,
Eq.(\ref{3_07}), of the Hamiltonian vanishes. In addition, the
condition $u^2_{Kk} + v^2_{Kk} = 1$ preserving Fermi's commutation
relations for quasiparticle operators must be fulfilled. These
conditions yield
\begin{eqnarray}
\label{3_09} && v^2_{Kk}={\frac 12}\left (1- \frac
{\xi_{Kk}^{}}{\sqrt
{{\xi^2_{Kk}}+{\Delta^2_{Kk}}}} \right ), \nonumber\\
&&  u_{Kk}^{}v_{kk}^{} = -{\frac 12}{\frac {\Delta_{Kk}^{}}{\sqrt
{\xi^2_{Kk}+\Delta^2_{Kk}}}}.
\end{eqnarray}
It should be noted that the Bogoliubov--Valatin amplitudes
(\ref{3_09}) are determined not by the one-particle energy
$\varepsilon ({\bm k}) - \mu$ as it were at ${\bm K}=0$ but by the
relative motion energy (\ref{2_01}) which transforms into
$\varepsilon ({\bm k})- \mu$ just in the case ${\bm K}=0$.

In the zero-temperature limit, one can obtain the self-consistency
equation determining the SC order parameter,
\begin{equation}
\label{3_10}\Delta_{Kk}^{}=-{\frac 1{2S}}\sum_{k^\prime}{\frac
{U({\bm k}-{\bm k^\prime})\Delta_{Kk^\prime}^{}}{\sqrt
{\xi^2_{Kk^\prime}+\Delta^2_{Kk^\prime}}}},
\end{equation}
where $\bm k$ and $\bm k'$ are momenta of the relative motion of
pairs with one and the same total momentum ${\bm K}$,
$U(\bm{\kappa})$ a matrix element of the effective interaction
potential in the Hamiltonian Eq.(\ref{3_01}) depending on the
momentum transfer $\bm {\kappa} = {\bm k}-{\bm k'}$ due to a
scattering from an initial ($\bm k$) to a finite ($\bm k'$) state
of the relative motion of the pair being also a difference, ${\bm
k^{}_{+}}-{\bm k^{\prime}_{+}}$, between the corresponding momenta
of the particles composing pairs with the total momentum ${\bm
K}$. Summation over momenta of the relative motion in the equation
(\ref{3_10}) is performed within the whole of their domain of
definition $\Xi_K^{}$. Thus, the SC order parameter is defined
inside the domain $\Xi_K^{}$ as well.

To solve the equation (\ref{3_10}) it is convenient to reduce it
to the corresponding integral equation considering the quantities
$\Delta_{\bm {Kk}}^{}=\Delta (\bm k)$ and $\xi_{\bm {Kk}}=\xi(\bm
k)$ as continuous functions of the relative motion momentum.
Hereafter, for a simplification of all of the notations, we omit
the label ${\bm K}$. Using the notations
\begin{equation}
\eta(\bm k)=\sqrt{\xi^2_{}(\bm k)+\Delta^2_{}(\bm k)}, \quad f(\bm
k)=1/{{4\pi}{\eta(\bm k)}}, \label{3_11}
\end{equation}
one can rewrite (\ref{3_10}) in the form
\begin{equation}
\Delta(\bm k)=-{\frac1{2\pi}}\int_{\Xi}U(|{\bm k}-{\bm k'}|)f(\bm
k')\Delta(\bm k')d^2_{}k', \label{3_12}
\end{equation}
where $U(|{\bm k}-{\bm k'}|)$ may be called a kernel of the
non-linear integral equation (\ref{3_12}).

\section{Necessary conditions of
repulsion--induced SC pairing}

In this section, we discuss the necessary conditions of the
repulsion-induced SC pairing with large total pair momentum and
put forward a procedure of reducing of the mean-field SC gap
equation (\ref{3_12}) to an approximate non-linear singular
integral equation with a degenerate kernel.\cite{BKSS}

The matrix element $U({\bm k}-{\bm k'})$ is connected with the
effective interaction potential $U(r)$ in the real space by the
Fourier transform
\begin{equation}
U(|{\bm k}-{\bm k'}|) = \int U(r)\exp{[i({\bm k}-{\bm k'}){\bm
r}]}d^2_{}r. \label{4_01}
\end{equation}
One can deduce a simple criterion of the absence of non-trivial
solutions of the equation (\ref{3_12}). Let us multiply this
equation and the function $f(\bm k)\Delta(\bm k)$ together and
integrate the obtained relation over ${\bm k}$ inside the domain
$\Xi$. Expressing $U({\bm k}-{\bm k'})$ in accordance with
(\ref{4_01}) and changing the order of integration over ${\bm k},
{\bm k'}$ and ${\bm r}$ one can obtain
\begin{equation}
\int_{\Xi}{\Delta}^2_{}({\bm k})f({\bm k})d^2_{}k = -
{\frac1{2{\pi}}}\int U(r)L({\bm r})d^2_{}r, \label{4_02}
\end{equation}
where
\begin{equation}
L({\bm r})=\left |\int_{\Xi}{\Delta}({\bm k})f({\bm k})\exp{(i{\bm
k}{\bm r})}d^2_{}k \right|^2 \label{4_03}
\end{equation}
is a nonnegative function of ${\bm r}$. The integration over ${\bm
r}$ in (\ref{4_02}) is performed over the whole of 2D real space.
As far as $f({\bm k})\geq 0$, the left side of (\ref{4_02}) is
nonnegative as well. If the potential $U(r)$ were positive at any
${\bm r}$, the right-hand side of the Eq. (\ref{4_02}), on the
contrary, turns out to be negative and the equality (\ref{4_02})
may be satisfied only in the trivial case when $\Delta ({\bm k})
\equiv 0$ in the whole of the domain $\Xi$. Thus, under the
condition that the interaction is purely repulsive in the real
space that is $U(r) >0$ at any $r$, the self-consistency equation
leads only to the trivial solution.

Therefore, a rise of repulsion-induced SC order is possible only
in the systems where the interaction potential is a function of
$r$ with alternating signs. As an example of such a potential with
alternating signs, one can consider screened repulsive Coulomb
interaction in a degenerated Fermi system which exhibits the
Friedel oscillations. \cite{Kittel} Further, we restrict
ourselves to a consideration of such a repulsive potential $U(r)$.

One can see that the maximal (positive) value of the function
$rU(r)$, arising when the integral in the right-hand side of
(\ref{4_02}) is calculated using the polar coordinates,
corresponds to $r=0$. This function with alternating signs is
multiplied by a nonnegative function $L({\bm r})$ defined by
Eq.(\ref{4_03}). The governing contribution into the product
$\Delta ({\bm k})f({\bm k})$ determining the function $L({\bm r})$
is due to a small vicinity of the PFC because of the equality $\xi
({\bm k}) =0$ which takes place just on the PFC.

As it will be shown below, the gap function $\Delta ({\bm k})$
resulting from the repulsion interaction changes sign on a line
crossing the PFC. Therefore, the absolute value of the integral in
(\ref{4_03}) represents a function of $r$ appearing as a series of
peaks with decreasing magnitude and a character separation between
the neighboring peaks of about ${\pi}/k^{}_P$ where $k^{}_P$ is of
about the length of the PFC. The first (and the largest) peak
results in the fact that the product of the factor $\Delta ({\bm
k})f({\bm k})$ with alternating signs and exponential function in
the integral (\ref{4_03}) turns out to be a function with the
values mainly of constant sign.

The value $L(0)$ is, generally speaking, a small quantity because
of a partial (or even full) compensation of the contributions of
the parts of the domain $\Xi$ in which $\Delta ({\bm k})$ has
opposite signs. For example, in the case of the full compensation,
$L(r)\sim r_{}^4$, therefore the main positive extremum of the
function $rU(r)$ turns out to be suppressed considerably. In the
opposite case of large $r$, the function $L(r)$ is also small due
to rapidly oscillating factor $\exp(i{\bm kr})$. As far as
$k^{}_P$ is pronouncedly less than $k^{}_F$, the position of the
first maximum of the function (\ref{4_03}), $r^{}_1 \sim
{\pi}/{k^{}_P} > {\pi}/{k^{}_F}$, corresponds quite naturally to
the region of the Friedel oscillations of the repulsive Coulomb
potential. There is a significant region of the $r$~-~space in
which the integrand in the right hand side of the Eq.(\ref{4_02})
is negative whereas the contributions of the regions with positive
integrand are comparatively small as it is illusyrated
schematically in Fig.2. Therefore the integral in (\ref{4_02}) may
turn out to be negative.

\begin{figure}
\includegraphics[]{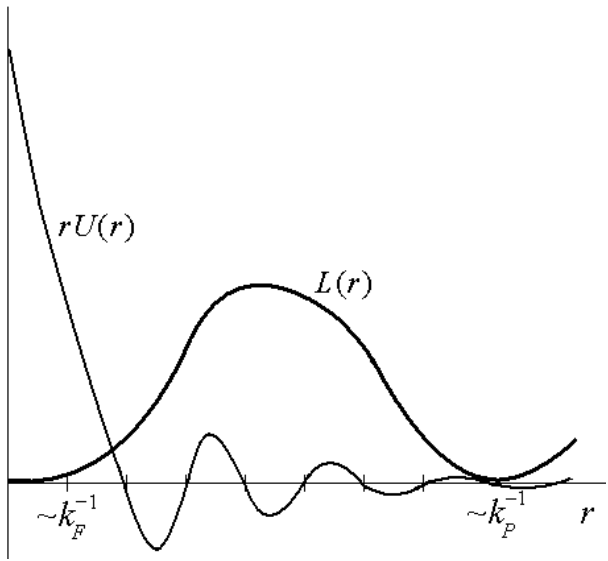}
\caption[*]{\label{bks_02.eps} Behavior of the functions $rU(r)$
and $L(r)$ contributing to the right-hand side   integral in
Eq.(\ref{4_02}), schematically. One can see $L(r)$ as the
function (\ref{4_03}) averaging over the angle variable.}
\end{figure}

In such a case, the equality (\ref{4_02}) is provided due to the
fact that the function $f({\bm k})$ is {\it{singular}} on the PFC
and may be {\it{arbitrarily large}} at $\Delta ({\bm k})
\rightarrow 0$. The magnitude of the oscillations of the function
$L(r)$ depends on $\Delta ({\bm k})$ because an effective width of
the vicinity of the PFC leading to the main contribution into the
integral (\ref{4_02}) is determined by just the function $\Delta
({\bm k})$. Thus, the condition (\ref{4_02}) may be fulfilled at
sufficiently small $\Delta ({\bm k}) \neq 0$.

Therefore {\it{the presence of the PFC}} which results in the
singularity of the Eq.(\ref{3_12}) is {\it{the first necessary
condition}} of the repulsion-induced SC pairing.

The equation (\ref{3_12}) is a non-linear integral Hammerstein
equation with a symmetric kernel $U(|{\bm k}-{\bm k'}|)$. All of
the eigenvalues of such a nondegenerate kernel are real and belong
to a discrete spectrum ${\lambda}^{}_n$ (where $n=1,2,\dots )$
with the condensation point $|{\lambda}^{}_n |\rightarrow \infty$
at $n \rightarrow \infty$. If a kernel $U(|{\bm k}-{\bm k'}|)$ is
the Fourier transform (\ref{4_01}) of everywhere positive
effective potential $U(r)$ (a positive defined kernel) all its
eigenvalues are positive as well. In such a case, as it follows
from Eq.(\ref{4_02}), the equation (\ref{3_12}) has the trivial
solution only.

Therefore, the presence of at least one {\it{negative eigenvalue}}
in the spectrum of the kernel $U(|{\bm k}-{\bm k'}|)$ is {\it{the
second necessary condition}} of the existence of a non-trivial
solution of the equation (\ref{3_12}).

It should be noted that, as it follows from Eq.(\ref{3_12}), a
non-trivial solution $\Delta ({\bm k})\neq 0$ of the
self-consistency equation at $U({\bm k}-{\bm k'})
>0$, if such a solution exists, {\it {must be a function
with alternating signs}} inside its domain of definition $\Xi$.

In the case of the non-degenerate kernel $U(|{\bm k}-{\bm
{k}^{\prime}}|)$, the existence of such a non-trivial solution of
the self-consistency equation at the repulsive interaction can be
demonstrated if one represents this kernel as an expansion over
its eigenfunctions ${\varphi}^{}_n({\bm k})$ which are the
solutions of the linear integral equation
\begin{equation}
{\varphi}^{}_n({\bm k})= {\lambda}^{}_n \int_{\Xi}U(|{\bm k}-{\bm
k_{}^{\prime}}|){\varphi}^{}_n({\bm k}_{}^{\prime})
d_{}^2k_{}^{\prime} . \label{4_04}
\end{equation}
Let us write this expansion in the form \cite{Mikhlin}
\begin{equation}
U({\bm k}-{\bm k'}) = \sum_n {\frac {{\varphi}^{\ast}_n({\bm
k}){\varphi}^{}_n({\bm k}_{}^{\prime})}{{\lambda}^{}_n }}
\label{4_05}
\end{equation}
and substitute it into the self-consistency equation (\ref{3_12}).
Integrating the right-hand side of this equation term-by-term, one
can see that a non-trivial solution, if it exists at all, may be
represented in the form of an expansion over the complete system
of the eigenfunctions ${\varphi}^{}_n({\bm k})$,
\begin{equation}
{\Delta}({\bm k}) = \sum_{n^{\prime}} {\Delta}^{}_{n^{\prime}_{}}
{\varphi}^{}_{n_{}^{\prime}}({\bm k}), \label{4_06}
\end{equation}
where ${\Delta}^{}_n$ are the expansion coefficients which can be
determined as the solutions of the infinite equation system,
\begin{equation}
{\Delta}^{}_n = - {\frac 1{2{\pi}{\lambda}^{}_n}}
\sum_{n^{\prime}} f^{}_{n n^{\prime}}{\Delta}^{}_{n^{\prime}_{}}.
\label{4_07}
\end{equation}
Here, $f^{}_{n n^{\prime}}$ are the matrix elements of the form
\begin{equation}
f^{}_{n n^{\prime}} = \int_{\Xi} {\varphi}^{\ast}_n({\bm k})
f({\bm k}) {\varphi}^{}_{n^{\prime}}({\bm k}) \label{4_08}
\end{equation}
where the function $f({\bm k})$ is determined by Eq.(\ref{3_11}).

Multiplying Eq.(\ref{4_07}) and ${\lambda}^{}_n {\Delta}^{}_n$
together, one can, after a summation over $n$, obtain the relation
\begin{equation}
\sum_n {\lambda}^{}_n {\Delta}^2_n = -{\frac 1{2 \pi}} \sum_{n,n'}
{\Delta}^{}_n  f^{}_{n n^{\prime}} {\Delta}^{}_{n'} \label{4_09}
\end{equation}
which is analogous to the relation (\ref{4_02}). As it is follows
from Eq.(\ref{4_09}), there is only the trivial solution of this
equation in the case when all of the eigenvalues are positive.
Therefore, a non-trivial solution may, in principle, exist only in
the case when at least one of the eigenvalues turns out to be
negative. In fact, such a condition turns out to be not only
necessary but the sufficient one.

To prove this statement, one can take advantage of the obvious
analogy between the spectral expansion of the order parameter,
Eq.(\ref{4_06}), and its expansion into a series of spherical
harmonics. As it is well known, the SC pairing arises when the
negative scattering length corresponds at least to one of the
values of orbital angular momentum. \cite{Lifshitz} For this
reason, taking account of such an analogy, we do not present a
direct proof of such a statement within the framework of the
spectral expansion.\cite{S}

\section{Simple degenerate kernel}

The function defining a non-linear operator in Eq.(\ref{3_12}) has
a special form ${\Delta} f(\Delta)$ where $f(\Delta)$ may be
called a nonlinearity factor. In this sense, the equation
(\ref{3_12}) may be referred to the class of quasi-linear integral
equations. In the case of a degenerate kernel, the significant
feature of such equations is that a form of a non-trivial solution
(if it exists at all) reproduces the kernel structure. The kernel
$U(|{\bm k}-{\bm k'}|)$ can be approximately reduced to a
degenerate one if one takes into account that both variables, $\bm
k$ and $\bm k'$, are defined in a relatively small domain $\Xi$ of
the momentum space. Therefore, if the function (\ref{4_01}) can be
expanded into the Tailor series, one can restrict himself to some
lowest powers of the argument $\kappa = |{\bm k}-{\bm k'}|$. As it
follows from the definition (\ref{4_01}), the expansion of
$U(\kappa)$ includes only even powers of $\kappa$ and the Fourier
transform of the interaction potential can be written as
\begin{equation}
U(\kappa) = 2\pi \left [ u^{}_0 - {\frac12}\;u^{}_2
\;{\kappa}^2_{} + {\frac {3}{8}}\; u^{}_4 \;{\kappa}^4_{} -
...\right ], \label{5_01}
\end{equation}
where
\begin{equation}
u^{}_n = {\frac 1{n!}}\; \int_0^{\infty}U(r)\;r_{}^{n+1}\;dr.
\label{5_02}
\end{equation}
A degenerate kernel corresponds to a finite number of the terms in
the expansion (\ref{5_01}) and may be considered as a good
approximation of the true nondegenerate kernel in the case of a
small domain $\Xi$. It should be noted that the Fourier transform
$U(\kappa )$ corresponding to a repulsive interaction is, at small
$\kappa$, a decreasing function of its argument resulting in
$u_0^{}>0$ and $u^{}_2 >0$.

To solve the equation (\ref{3_12}) it is convenient to transform
it into a dimensionless form. Using two constants $U_0^{}$ and
$r_0^{}$, being characteristic energy and length scales,
respectively, one can define them by the relations $U_0^{}r_0^2=
u^{}_0$ and $U_0^{}r_0^4= u^{}_2$. In the case of screened Coulomb
potential the parameters $r_0^{}$ and $U_0^{}=e^2_{}/r_0^{}$ make
sense of a screening length and a characteristic Coulomb energy
respectively. Thus, the energy $U_0^{}$ determines the scale of
the quantities $\Delta(\bm k)$ and $\xi(\bm k)$ whereas a momentum
is measured by the units of $r_0^{-1}$. The equation (\ref{3_12})
remains invariant with respect to such a scaling if one reduces
the domain of integration $\Xi$ to dimensionless variables (the
components $k_1^{}$ and $k_2^{}$ of the momentum of the relative
motion of the pair).

If the domain $\Xi$ is small enough one may keep only two terms of
the lowest order in the expansion (\ref{5_01}) and obtain a
degenerate kernel in the form
\begin{equation}
U^{}_d(\kappa) = 2\pi [1-{{\kappa}_{}^2}/2]. \label{5_03}
\end{equation}
One can make sure that the obtained degenerate kernel has four
eigenvalues three of which are positive whereas the fourth one is
negative.

To show this one has to write down the linear integral equation
determining the eigenvalues $\lambda$ and the eigenfunctions
${\varphi}^{}_{\lambda}({\bm k})$ corresponding to the degenerate
kernel Eq.(\ref{5_03}),
\begin{equation}
{\varphi}^{}_{\lambda}({\bm k})= {\lambda} \int_{\Xi}U^{}_d(|{\bm
k}-{\bm k_{}^{\prime}}|){\varphi}^{}_{\lambda}({\bm
k}_{}^{\prime}) d_{}^2k_{}^{\prime} . \label{5_04}
\end{equation}
It follows immediately from the Eq.(\ref{5_04}) and the form of
the kernel (\ref{5_03}) that eigenfunctions must be taken in the
form
\begin{equation}
{\varphi}^{}_{\lambda}({\bm k})= a+({\bm {\chi}k})+bk_{}^2
\label{5_05}
\end{equation}
thus reproducing the form of the degenerate kernel (\ref{5_03}).
Here, $a$ and $b$ are scalar coefficients and ${\bm {\chi}}$ is a
constant vector being subjects to be determined for each of the
eigenvalues. To obtain the parameters $a$, $b$ and ${\bm {\chi}}$
of the solution (\ref{5_05}) one has to substitute this solution
unto the Eq.(\ref{5_04}) and, after integration, set the
coefficients corresponding to one and the same power of the vector
${\bm k}$ equal to each other.

It is obvious that the approximation (\ref{5_03}) is sufficient if
$(k+k')^2_{} <2$. Therefore, one has to assume that the domain
$\Xi$ is such that the condition
\begin{equation}
k^2_{}<1/2. \label{5_06}
\end{equation}
is fulfilled for any momentum of the relative motion. If one takes
into account that a screening length $r_0^{}$ is of about a few
interatomic distances, the restriction (\ref{5_06}) results in the
fact that a characteristic size of the domain $\Xi$ should be much
less as compared with a characteristic Brillouin zone size.

Due to a symmetry of the domain of integration $\Xi$ with respect
to inversion transformation ${\bm k} \leftrightarrow - {\bm k}$
eigenfunctions ${\varphi}^{}_{\lambda}({\bm k})$ must be either
even or odd. Odd eigenfunctions are determined by single unknown
vector ${\bm {\chi}}$ being the solution of the equation
\begin{equation}
{\bm {\chi}}=2{\pi} {\lambda} \int_{\Xi}({\bm {\chi}}{\bm
k}_{}^{\prime}){\bm k_{}^{\prime}}d_{}^2k_{}^{\prime}.
\label{5_07}
\end{equation}
This equation has two non-trivial solutions directed along the
symmetry axes of the domain $\Xi$. Corresponding eigenvalues,
\begin{equation}
{\lambda}^{}_i =\left[2{\pi} \int_{\Xi} k^2_i d_{}^2k \right
]_{}^{-1},\quad i=1,2 \;, \label{5_08}
\end{equation}
are positive. Here, $k^{}_i$ are the components of the vector
${\bm k}$ along the symmetry axes of the domain $\Xi$.

In the case of the even eigenfunctions, one can obtain the system
of linear homogeneous equations
\begin{eqnarray}
a&=&2{\pi}{\lambda}\int_{\Xi}(1-{k_{}^{\prime}}_{}^2/2)
(a+b{k_{}^{\prime}}_{}^2)d_{}^2k_{}^{\prime}, \nonumber \\
b&=&{\pi}{\lambda}\int_{\Xi}
(a+b{k_{}^{\prime}}_{}^2)d_{}^2k_{}^{\prime} \label{5_09}
\end{eqnarray}
determining two unknown quantities, $a$ and $b$. The condition of
the non-trivial compatibility of this system results in another
two eigenvalues of the degenerate kernel Eq.(\ref{5_03}),
\begin{equation}
{\lambda}^{}_{\pm} ={\frac 1{\pi}}\left[(K^{}_0 -K^{}_1)\pm
\sqrt{(K^{}_0-K^{}_1)_{}^2 +(K^{}_0K^{}_2-K^2_1)}\right
]_{}^{-1}.\label{5_10}
\end{equation}
Here we use the notation
\begin{equation}
K^{}_n =\int_{\Xi}k_{}^{2n}d_{}^2k, \quad n=0,1,2.\label{5_11}
\end{equation}

As it follows from the Cauchy--Schwarz--Bunyakovsky inequality for
the integrals (\ref{5_11}), $K^{}_0K^{}_2\geq K^2_1$, therefore
one of the eigenvalues (\ref{5_10}) is positive, ${\lambda}^{}_+
>0$, whereas the second one is negative, ${\lambda}^{}_- <0$. Here
we take into account the fact that, due to the inequality
(\ref{5_06}), $2K^{}_1<K^{}_0$.

It should be noted that two of the positive eigenvalues
${\lambda}^{}_{1,2}\sim u_0^{-1}$, another one positive eigenvalue
${\lambda}^{}_+ \sim u_2^{-1}$, and the negative eigenvalue
${\lambda}^{}_- \sim - u_0^{}/u_2^2$.

One may neglect the next terms of the expansion (\ref{5_01}) only
under the condition that, inside the domain $\Xi$, the mean square
of the neglected terms is much less in comparison with the
absolute value of any eigenvalue of the degenerate kernel:
\cite{Mikhlin}
\begin{equation}
2|u^{}_4| \sqrt{\int_{\Xi} d^2_{}k \int_{\Xi} d^2_{} k' ({\bm
k}-{\bm k'})^8_{}} \ll {|\lambda |}_{}^{-1}.\label{5_12}
\end{equation}
In the case of the positive eigenvalues, this inequality is
satisfied under rather natural condition that the third term in
the expansion (\ref{5_01}) is small in comparison with the second
one. In the case of the negative eigenvalue, the inequality
(\ref{5_12}) reduces to the condition that $u^{}_0 |u^{}_4 | < c
u^2_2$ where $c\ll 1$. In the case of everywhere positive
potential $U(r)$, this relation does not fulfilled definitely
since, for such a potential, the quantities $u^{}_n$ defined in
(\ref{5_02}) should be connected by a reciprocal inequality
following from the Cauchy--Schwarz--Bunyakovsky inequality for the
integrals (\ref{5_02}). Thus, using the degenerate kernel
(\ref{5_03}) instead of the true kernel $U({\bm k}-{\bm k'})$, one
must assume that the conditions (\ref{5_06}) and (\ref{5_12}) are
fulfilled.

\section{General solution of the self-consistency equation}

Substituting the kernel (\ref{5_03}) into the equation
(\ref{3_12}) and grouping the terms independent of the momentum
and also linear and squared terms one can see that the dependence
of the order parameter on the momentum of the relative motion of
the pair has the form similar to the Eq.(\ref{5_05})
\begin{equation}
\Delta({\bm k})=a + ({\bm {\chi k}}) -b k^2_{}, \label{6_01}
\end{equation}
where $a$ and $b$ are unknown parameters and ${\bm \chi}$ is  an
unknown constant vector being subjects to be determined. Thus, one
can see that the solution of the equation (\ref{3_12}) reproduces
the momentum dependence of the degenerate kernel. Therefore, to
solve this equation, one has to determine the parameters $a$, $b$
and ${\bm \chi}$. To find these parameters one has to substitute
the kernel (\ref{5_03}) and an explicit form of the solution
(\ref{6_01}) into the equation (\ref{3_12}). After that, one
obtains a system of integral equations determining the parameters
$a$, $b$ and ${\bm \chi}$. One of these equations, following from
the comparison of the coefficients at the first power of $\bm k$
in the left and right sides of the equation resulting from the
Eq.(\ref{3_12}), has the form
\begin{equation}
{\bm \chi}=-\int_{\Xi} f(\bm k)\Delta({\bm k}){\bm k}d^2_{}k.
\label{6_02}
\end{equation}

The equation (\ref{6_02}) determines the vector ${\bm \chi}$
implicitly because the right-hand side of this equation depends on
$\Delta$ which itself depends on ${\bm \chi}$. It is not difficult
to see that there is at most one solution of the equation
(\ref{6_02}). To prove this proposition let us introduce an
auxiliary function $H(\Delta)\equiv \Delta \cdot f(\eta(\Delta))$
which is a monotonically increasing function of its argument.
Writing down the equation (\ref{6_02}) in the component-wise form
one can see that its right-hand side is a decreasing function of a
component $\chi_i^{}$ of the vector ${\bm \chi}$ ($i=1,2$) whereas
the left-hand side increases with $\chi_i^{}$. Therefore the
equality (\ref{6_02}) turns out to be possible at a single value
of the vector parameter ${\bm \chi}$.

The obvious solution of (\ref{6_02}) is ${\bm \chi}=0$. Indeed,
due to central symmetry of the domain $\Xi$, the right-hand side
of (\ref{6_02}) vanishes at ${\bm \chi}=0$ as a result of an
integration of an odd function over a symmetric domain. It should
be noted that the order parameter $\Delta ({\bm k})$ and the
excitation energy $\xi(-{\bm k})=\xi({\bm k})$ are even functions
of ${\bm k}$ at ${\bm \chi}=0$. There are no other solutions of
the equation (\ref{6_02}). Thus, in the case of a degenerate
kernel (\ref{5_03}), the equation (\ref{3_12}) may presuppose a
simple solution of the form
\begin{equation}
\Delta(k)=a - b k^2_{}, \label{6_03}
\end{equation}
determining the SC gap as a parabolic function of the momentum of
the relative motion of the pair.

Taking into account that ${\bm \chi}=0$ one can write down two
equations determining the parameters $a$ and $b$:
\begin{eqnarray}
\left(J_0^{}-{\frac12}J_1^{}+1\right)a-
\left(J_1^{}-{\frac12}J_2^{}\right)b=0, \nonumber \\
{\frac12}J_0^{}a -\left({\frac12}J_1^{}-1\right)b=0, \label{6_04}
\end{eqnarray}
where
\begin{equation}
J_n^{}\equiv \int_{\Xi} f({\bm k}) k^{2n}_{}d^2_{}k, \label{6_05}
\end{equation}
where $n=0,1,2$. Thus, we have a quasi-linear system of the
equations (\ref{6_04}) which contains three positive integrals
(\ref{6_05}) depending on two unknown parameters $a$ and $b$.

First of all, let us multiply the inequality $k^2_{}<1/2$ by the
function $f(\eta ({\bm k}))$ and then by the another function,
$k^2_{}f(\eta ({\bm k}))$. Integrating the obtained inequalities
over the domain $\Xi$ one can deduce two new inequalities,
$J_1^{}<J_0^{}/2$ and $J_2^{}<J_1^{}/2$, which allow to conclude
that both coefficients in the first equation (\ref{6_04}) are
positive. Therefore, the solutions of the equation system
(\ref{6_04}) have one and the same sign (positive, for example).
As a result, the SC order parameter as a function of the momentum
of the relative motion changes its sign at a certain $k=k_0^{}$
and can be written as
\begin{equation}
\Delta (k) = b (k_0^2 -k_{}^2) \label{6_06}
\end{equation}
where $k^2_0=a/b$. The parameter $b$ determines an energy scale of
the SC gap and $k_0^{}$ is a radius of the circle on which the SC
gap vanishes. In the case of repulsive interaction considered
here, it is obvious that the SC gap has to vanish inside the
domain $\Xi$ because at $U(|{\bm k}-{\bm k'}|)>0$ there exists no
constant-sign non-trivial solution of the equation (\ref{3_12}).

The conditions that $a >0$ and $b >0$, as it follows from the
second equation (\ref{6_04}), lead to a restrictive inequality
$J_1^{}>2$ which is a consequence of more strong inequality
following from the condition of the non-trivial consistency of the
system (\ref{6_04}). Calculating the determinant of the system
(\ref{6_04}) one can write down this condition in the form
\begin{equation}
(J_1^{}/2 - 1)^2_{} = J_0^{} (J_2^{}/4 - 1). \label{6_07}
\end{equation}
It is seen that it must be $J_2^{}>4$. Therefore, taking into
account that $J_1^{}>2J_2^{}$, one can obtain $J_1^{}>8$.

Instead of the pair parameters $a$ and $b$ characterizing the SC
gap it is convenient to consider another pair, $b$ and $k^{}_0$.
To determine these two parameters one can use any pair of the
equations (\ref{6_04}) and (\ref{6_07}). As one of them it is
convenient to use the second equation of the system (\ref{6_04})
writing it in the form
\begin{equation}
J_1^{} - k_0^2J_0^{} =2. \label{6_08}
\end{equation}
Then, expressing the left-hand side of the equation (\ref{6_07})
with the help of the second of the equations (\ref{6_04}), the
Eq.(\ref{6_07}) can be rewritten as
\begin{equation}
J_2^{} - k_0^4J_0^{} =4. \label{6_09}
\end{equation}
The system of equations (\ref{6_08}) and (\ref{6_09}) is fully
equivalent to initial system (\ref{6_04}) and can be used to
determine the parameters $b$ and $k^{}_0$ of the SC gap. In this
connection, it should be noted that the integrals $J_0^{}$,
$J_1^{}$ and $J_2^{}$, because of their dependence on $\Delta $,
depend themselves on $b$ and $k^{}_0$.

\section{order parameter}

Let us rewrite the equations (\ref{6_08}) and (\ref{6_09}) in the
explicit form
\begin{equation}
{\frac 1{4\pi}} \int _{\Xi} {\frac {(k_{}^{2n}-k_0^{2n}) d^2_{}k
}{\sqrt{\xi^2_{}({\bm k})+{b} ^2_{}(k_{}^2-k_0^2)^2_{}}}}=2_{}^n,
\label{7_01}
\end{equation}
where $n=1,2$. Is is impossible to solve the equation system
(\ref{7_01}) analytically in the whole of the domain of definition
of unknown quantities ($b>0,\;0<k_0^2 < 1/2$) because the
integrals in (\ref{7_01}) depend on the form of the domain of
integration. However, it is easy to determine these integrals in
the two limiting cases of large ($b \rightarrow \infty$) and small
($b \rightarrow +0$) values of the gap energy scale $b$. Indeed,
at large $b$ one can neglect the augent in the radicand of each of
the integrals (\ref{7_01}). After that, these integrals can be
calculated easily and turn out to be proportional to $1/b$
therefore one can obtain the solution of (\ref{7_01}) in a trivial
way.

Considering the integrals in (\ref{7_01}) in the most interesting
case of small $b$ one has to take into account the fact that pair
excitation energy (\ref{2_01}) vanishes on the line separating
occupied, $\Xi_{}^{(-)}$, and unoccupied, $\Xi_{}^{(+)}$, parts of
the domain $\Xi$ as far as this line (PFC) represents a piece of
the FC. This results in a divergence of the integrals (\ref{6_05})
at small $b$ and, if one takes into account that the main
contribution into the integrals is from a narrow strip along the
line $\xi ({\bm k})=0$, it is this divergence that results in the
existence of the solution of the system (\ref{7_01}). Let us
denote a width of the strip as $2 \Delta k$ and, taking account of
the equation of the line $\xi ({\bm k})=0$ in polar coordinates,
$k=k(\varphi)$, consider an integral of a general form
\begin{equation}
J={\frac 1{4\pi}} \int _{\Xi}{\frac {F(k)\;k\; dk\;d\varphi}{\sqrt
{\xi^2_{}(k,\varphi)+ b ^2_{}(k_0^2 -k^2_{})^2_{}}}}, \label{7_02}
\end{equation}
where $F(k)$ is a continuous function without any singularity. Let
us consider a behavior of this integral at $b \rightarrow +0$
using a strip of the width $2 \Delta k$ as an integration domain.
First of all, one can perform an integration over a polar angle
in the limits from $\varphi _1^{}$ up to $\varphi _2^{}$
corresponding to the endpoints of the line $\xi (k, \varphi) = 0$.
Then, introducing a new variable of integration $\xi$, one can
perform an integration over $k$ in the limits from $k(\varphi ) -
\Delta k$ up to $k(\varphi ) + \Delta k$. On account of the
smallness of $\Delta k$, one can assume that $k$ and $\xi$ are
connected with each other in a linear way: $\xi \simeq (d \xi
(k(\varphi ))/dk)(k-k(\varphi ))$. In addition, the argument $k$
in the integrand can be replaced by its constant (at a given
$\varphi$) value corresponding to the PFC: $k=k(\varphi )$. Thus,
after integration over $\xi$, one can obtain a singular
(logarithmic in $b$) contribution into the integral (\ref{7_02}).

Besides the singular contribution there is also a regular one
which depends on one-particle dispersion and on the form and size
of the domain $\Xi$. This contribution depends on $k^2_0$ as on a
parameter. Thus, the integral (\ref{7_02}) can be represented as
\begin{equation}
J= A \ln (1/b) + C, \label{7_03}
\end{equation}
where the coefficient in front of the logarithm has the form
\begin{equation}
A={\frac 1{2\pi}} \int _{\varphi _1^{}}^{\varphi _2^{}}{\frac{k
F(k)}{|\xi '_k |}} d\varphi, \label{7_04}
\end{equation}
where an integration over the angle variable should be performed
along the PFC (see Fig.1). The regular part of the integral
denoted as $C$ can be taken equal to its value at $b=0$.

Thus, at small $b$, the equation system (\ref{7_01}) can be
written in the form
\begin{equation}
(A_n^{}-k_0^{2n} A_0^{})\ln(1/b)=P^{}_n, \label{7_05}
\end{equation}
where the coefficients $A_n^{}$ are independent of $k^2_0$ and
can be determined by the expression (\ref{7_04}) in which one has
to put $F(k)=k_{}^{2n}$ whereas the right-hand sides of the
equations (\ref{7_05}) depend on $k^2_0$ and can be written as
\begin{equation}
P_n^{}=2^n_{}-C_n^{}+k_0^{2n}\;C_0^{}.\label{7_06}
\end{equation}
Here $C^{}_n$ are the regular parts of the integrals (\ref{6_05}).

As it follows from the Eqs.(\ref{7_05}), the equation system
(\ref{7_01}) has a non-trivial solution with necessity. To prove
such a statement let us consider each of the equations
(\ref{7_01}) as an equation of a line on the plane $b ,\; k^2_0$:
$b^{}_n = b^{}_n (k^2_0)$. Then the point of the intersection of
these two lines corresponds to the solution of the equation
system (\ref{7_01}). Putting $k^2_0=0$ in these equations one can
obtain the left-hand sides of each of them as functions of $b$
(Fig.3). These functions monotonically decrease from $+\infty$ at
$b \rightarrow +0$ to zero at $b \rightarrow + \infty$. Due to the
inequality (\ref{5_06}), the second of these functions,
corresponding to $n=2$, with necessity is less than the first one
($n=1$) at any $b$. Therefore, at $k^2_0=0$, the solution
$b_1^{}(0)$ of the first equation turns out to be more than the
solution $b_2^{}(0)$ of the second equation because of the fact
that the right-hand side of the second equation is always more as
compared to the right-hand side of the first one. In the another
limiting case when $b \rightarrow +0$, one can use an asymptotic
representation of the equations (\ref{7_01}) in the form
(\ref{7_05}). As far as the right-hand sides of the equations
(\ref{7_05}) are finite and $\ln {(1/b)} \rightarrow +\infty$ at
$b \rightarrow +0$ the values of $k^2_0$ corresponding to $b
\rightarrow 0$ can be written as $k^2_{01} =A^{}_1/A^{}_0$ at
$n=1$ and $k^{}_{02} =\sqrt {A^{}_2/A^{}_0} $ at $n=2$. Taking
into account the condition (\ref{5_06}) one can see that
$k^2_{0n} < 1/2$. There is Cauchy--Schwarz--Bunyakovsky
inequality, $A^2_1 \leq A^{}_0 A^{}_2$,  for the quantities
$A^{}_0$, $A^{}_1$ and $A^{}_2$ defining by such integrals as
(\ref{7_04}) therefore $k^2_{01} \leq k^2_{02}$ and, with
necessity,  there exists a point of the intersection of the lines
$b^{}_n = b^{}_n (k^2_0)$.

\begin{figure}
\includegraphics[]{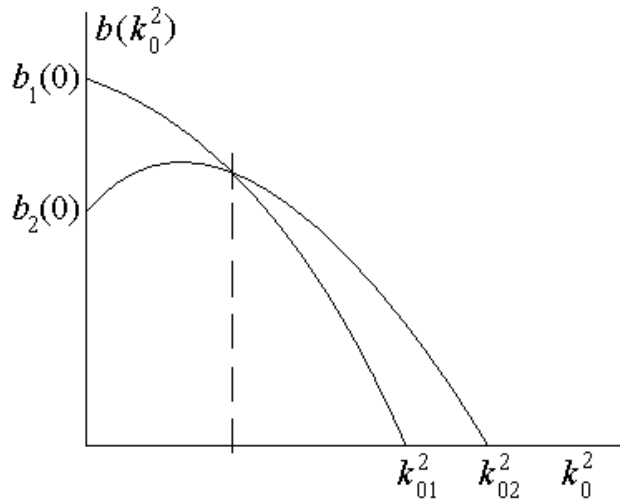}
\caption[*]{\label{bks_03.eps} Schematical illustration of the
graphic solution of the equation system (\ref{7_05}).}
\end{figure}

To obtain the non-trivial solution in the case of small $b$ one
has to perform term-by-term division of the equations
(\ref{7_05}). This leads to a closed equation determining the
parameter $k^2_0$. As a result, a characteristic energy scale of
the SC gap can be found from any equation of the system
(\ref{7_05}) and, for example, can be written in the form
\begin{equation}
b=\exp \left(-{\frac{P^{}_1}{A_1^{}-A_0^{}k_0^2}}\right).
\label{7_07}
\end{equation}
This relation explicitly determines the dependence of SC order
parameter on the momentum of the relative motion of the pair. Due
to the fact that $b <<1$, it is obvious that the exponent in
(\ref{7_07}) is negative.

One can see that, in the case of small $b$, the line $k=k_0^{}$
intersects the PFC. Indeed, let us suppose that $b$ and $k^{}_0$
are the solutions of the equation system (\ref{7_01}). At $b
\rightarrow +0$ the integrals in (\ref{7_01}) are determined by
their singular parts following from the integration over a small
vicinity of the PFC. An assumption that the circle $k=k^{}_0$
does not intersect the PFC means that both the PFC and its small
vicinity are situated either inside or outside this circle.
Therefore in the whole of this vicinity one may expect either the
inequality $k < k^{}_0$ or the opposite inequality $k
> k^{}_0$. In such a case, the integrand in (\ref{7_01}) must be
of constant sign and the integral corresponding to $n=2$ has to
be less than the integral with $n=1$ because of the inequality
$k^2_{} +k^2_0 < 1$ following from (\ref{5_06}). Thus, if one
consider that the circle $k=k^{}_0$ and the PFC does not
intersect each other the equations in the system (\ref{7_01})
turn out to be incompatible with each other. Consequently, at $b
\rightarrow +0$, the parameter $k^{}_0$ turns out to be such that
the intersection of the PFC and the circle $k=k^{}_0$ arises with
necessity. In such a case, the quantity $k^2_{} - k^2_0$ has
different signs on different pieces of the PFC interior and
exterior the circle $k=k^{}_0$. Such a conclusion proved under
the condition that $b \rightarrow +0$ valid also at rather small
but finite values of the parameter $b$.

As far as the circle $k=k^{}_0$ intersects the PFC, one can
assume that there exist the minimal, $k^{}_m$, and maximal,
$k^{}_M$, values of $k^{}_0$ corresponding to the endpoints of
the PFC. Taking into account the explicit expressions of the
integrals (\ref{7_04}) at $F(k)=k_{}^{2n}$, one can obtain an
estimation
\begin{equation}
|A^{}_1 -k^2_0 A^{}_0|\leq {\frac{\Delta \varphi}{2\pi}} {\frac
{k^3_M -k^3_m}{\langle v^{}_F \rangle }} \label{7_07'}
\end{equation}
where ${\Delta} {\varphi} = {\varphi}^{}_2 - {\varphi}^{}_1$ is an
angular size of the PFC and
\begin{equation}
{\frac 1{v^{}_F}}={\frac 1{{\Delta}
{\varphi}}}\int_{{\varphi}^{}_1} ^{{\varphi}^{}_2} {\frac {d\,
\varphi}{|{\xi}_{k}^{\prime}|}} \label{7_07''}
\end{equation}
is an average (over the PFC) value of the dimensionless Fermi
velocity. One can see from Eq.(\ref{7_07'}) that the absolute
value of the SC order parameter decreases exponentially with a
decrease both of the angular size and the anisotropy
(characterizing by the difference $k^{}_M -k^{}_m$) of the PFC.
In particular, in the case of repulsion-induced pairing with zero
total pair momentum (when the PFC coincides with the FC), the
non-trivial solution similar to Eq.(\ref{7_07}) is absent.

Thus, using conventional energy and momentum dimensions, one can
write down the SC order parameter, being the non-trivial solution
of the self-consistency equation (\ref{3_12}), in the form
\begin{equation}
{\Delta}(k)={\Delta}^{}_0(k) \exp{\left(-\frac1w \right)}
\label{7_08}
\end{equation}
where
\begin{equation}
w=(A^{}_1-A^{}_0k^2_0)/P^{}_1 \label{7_09}
\end{equation}
can be treated as an effective interaction strength and
\begin{equation}
{\Delta}^{}_0(k) =U^{}_0 r^2_0 (k^2_0-k^2_{}) \label{7_10}
\end{equation}
is a pre-exponential factor dependent on the momentum of the
relative motion of the pair.

\section{Step-wise approximation of the order parameter}

The continuous solution Eq.(\ref{7_08}) of the self-consistency
equation can be approximately presented as a step-wise function
of the momentum of the relative motion. \cite{BK_1,BKSS} Here, we
briefly discuss the simplest form of such an averaged solution
assuming that the true solution, Eq.(\ref{7_08}), is replaced by
an approximate one which has the form ${\Delta}(k)={\Delta}^{}_p$,
$p=1,2$, where
\begin{equation}
{\Delta}^{}_p =
{\frac{1}{{\Xi}^{}_p}}\int_{{\Xi}^{}_p}{\Delta}(k)d^2_{}k.
\label{8_01}
\end{equation}
The parameters ${\Delta}^{}_1$ and ${\Delta}^{}_2$ defined by the
Eq.(\ref{8_01}) may be considered as average values of the order
parameter inside the domain $\Xi$ at $k<k^{}_0$ and $k>k^{}_0$
respectively.

Performing the averaging of the left- and right-hand sides of the
self-consistency equation (\ref{3_12}) one can obtain the
equation system \cite{BKSS}
\begin{eqnarray}
2{\Delta}^{}_1 &=& - {\alpha}{\Xi}U^{}_{11}{\Delta}^{}_1
f^{}_1 - (1-{\alpha}){\Xi}U^{}_{12}{\Delta}^{}_2 f^{}_2, \nonumber\\
2{\Delta}^{}_2 &=& - {\alpha}{\Xi}U^{}_{21}{\Delta}^{}_1 f^{}_1 -
(1-{\alpha}){\Xi}U^{}_{22}{\Delta}^{}_2 f^{}_2. \label{8_02}
\end{eqnarray}
where $\alpha ={{\Xi}^{}_1}/{\Xi}$, $1-\alpha
={{\Xi}^{}_2}/{\Xi}$, and
\begin{equation}
f^{}_p \equiv f^{}_p ({\Delta }^{}_p)= {\frac
1{(2{\pi})^2_{}{{\Xi}^{}_p}}}\int_{{\Xi}^{}_p}{\frac
{d^2_{}k}{\sqrt{{\xi^2_{}({\bm k})+{\Delta^2_{p}}}}}} \label{8_03}
\end{equation}
is the value of the functional,
\begin{equation}
f^{}_p \{{\Delta }({\bm k})\}= {\frac 1{(2{\pi})^2_{}{\Delta
}^{}_p{{\Xi}^{}_p}}}\int_{{\Xi}^{}_p}{\frac {{\Delta({\bm
k})}d^2_{}k}{\sqrt{{\xi^2_{}({\bm k})+{\Delta^2_{}({\bm k})}}}}},
\label{8_04}
\end{equation}
corresponding to the average value ${\Delta}^{}_p$ of the order
parameter inside the subdomain ${\Xi}^{}_p$. The average value of
the interaction matrix element $U(|{\bm k}-{\bm k'}|)$
corresponding to a momentum transfer ${\bm {\kappa}}={\bm k}-{\bm
k'}$ is defined as
\begin{equation}
U^{}_{pp'}=U^{}_{p'p} ={\frac {1}{{{\Xi}^{\ast}_{pp'}}}}
\int_{{\Xi}^{\ast}_{pp'}}U(\kappa)d^2_{}{\kappa} \label{8_05}
\end{equation}
where ${{\Xi}^{\ast}_{pp'}}$ is a domain of definition of the
momentum transfer ${\bm {\kappa}}$ when an initial momentum,
${\bm k}$, belongs to the subdomain ${\Xi}^{}_p$ whereas a final
one, ${\bm k'}$, belongs to ${\Xi}^{}_{p'}$.

If one represents the functions (\ref{8_03}) in the explicit
form, the equations (\ref{8_02}) lead to a closed system of two
transcendental equations. The integrals (\ref{8_03}) over the
subdomains ${\Xi}^{}_p$ are similar to the integrals
(\ref{6_05}). Therefore, there exists a logarithmic singularity
of such an integral when the domain of integration includes the
PFC. The analysis of the non-trivial consistency of the
Eqs.(\ref{8_02}) shows that the non-trivial solution exists under
the condition that
\begin{equation}
U^{}_{12}U^{}_{21}-U^{}_{11}U^{}_{22}>0. \label{8_06}
\end{equation}
This condition coincides with the well-known
Suhl--Matthias--Walker condition \cite{SMW} and is definitely
fulfilled in the case of repulsive potentials which have an least
one negative eigenvalue.

The approximate approach developed here to determine the order
parameter arising at singlet repulsion-induced pairing and
corresponding to extended $s$~-wave symmetry can be used in the
case of $d$~-wave symmetry as well. In this connection, it should
be noted that a simple replacement \cite{Aoki} of the true kernel
$U({\bm k}-{\bm {k'}})$ of the self-consistency equation by a
constant
\begin{equation}
U({\bm k}-{\bm {k'}}) \rightarrow \langle {\Delta}({\bm k})
U({\bm k}-{\bm {k'}}){\Delta}({\bm {k'}})\rangle /\langle
{\Delta}^2_{}({\bm k})\rangle \label{8_07}
\end{equation}
treated as an average over the FC hardly ever leads to a
non-trivial solution because of the fact that such an averaging
procedure, similar to performed in Eqs.(\ref{4_02}) and
(\ref{4_09}), means the summation over both ${\bm k}$ and ${\bm {
k'}}$ resulting in a non-negative value of the constant
(\ref{8_07}) in the case of repulsive interaction. Thus, strictly
speaking, considering the scattering due to a repulsive
interaction $U({\bm k}-{\bm {k'}})$, one has to take into account
not only a difference between the areas of the momentum space
where the order parameter has different signs but also the
dependence of the interaction matrix element on the momentum
transfer ${\bm k}-{\bm {k'}}$.

\section{Conclusion}

In this paper, we have shown that there must be at least one
negative eigenvalue of the interaction matrix $U({\bm k}-{\bm
k'})$ to ensure the possibility of a repulsion-induced
non-trivial solution of the self-consistency equation. A rise of
the pair Fermi contour results in the fact that such a necessary
condition of repulsion-induced superconducting pairing becomes
the sufficient one at arbitrarily weak interaction strength.

It should be emphasized that negative eigenvalues have to arise
with necessity in real Fermi systems, in particular, due to
Friedel oscillations of screened Coulomb potential.
Quasi-two-dimensional electron system typical of cuprate
superconductors may be considered as an intermediate case between
comparatively weak Friedel oscillations in three-dimensional
system and one-dimensional charge density wave behavior. The pair
Fermi contour arises when, at certain large total pair momentum,
there exists mirror nesting of the Fermi contour.

Repulsion-induced singlet pairing results in an axtended $s$-wave
symmetry of the superconducting order parameter. This parameter
is defined inside the domain of definition of pair relative
motion momentum and changes its sign on a line intersecting the
pair Fermi contour. An anisotropy is a necessary feature of
one-electron dispersion which leads to the repulsion-induced
non-trivial solution of the self-consistency equation: the
corresponding superconducting order is absent in the framework of
any isotropic model.

\begin{acknowledgments}

The work was supported, in part, by the Russian
scientific-educational programme ``Integration'' (project B0049),
the Department of Education of Russian Federation (Grant
E02-3.4-147), the Russian Foundation for Basic Research (grant
02-02-17133) and Federal purposive technical-scientific programme
(State contracts N~40.072.1.1.1173, N~40.012.1.1.1357).
\end{acknowledgments}

\end{document}